# Polarization State Dynamics of Single Photon Pulse Under Stochastic Polarization Mode Dispersion for Optical Fiber Quantum Channels [*]

ZHU Chang-hua, PEI Chang-xing, QUAN Dong-xiao, CHEN Nan, YI Yun-hui

(State Key Laboratory of Integrated Services Networks, Xidian University, Xi'an 710071, China)

We investigate the polarization state dynamics of single photon pulse for optical fiber quantum communication channels. On the basis of a birefringence vector model in which amplitude and direction are both stochastic variables, Jone's vector is obtained by solving the frequency domain wave equation. The fidelity of output quantum state and degree of polarization of the pulse are also obtained from the density operators. It is shown that the fidelity of quantum state decreases quickly and tends to a stable value along optical fiber, and increases for larger mean fluctuation magnitude of the stochastic fiber birefringence. Degree of polarization is nearly constant for small mean fluctuation magnitude of the birefringence. The fidelity and degree of polarization vary in the same way for Gaussian and rectangular frequency spectrum envelope, while the value of Lorentzian spectrum is smaller.

In various practical quantum key distribution (QKD) systems, polarization coding and phase coding are mainly used [1]-[6]. The birefringence derived from inherent asymmetries of fiber core, mechanical stress, bending, and external field etc. has strong effect on the polarization state in the optical fiber quantum channel. The effect of devices in current fiber communications network, e.g. wavelength division multiplexer, on photon pulse polarization state can not be ignored [7]. Polarization controller might be used in practical QKD system for compensating the polarization state dynamically [8]. So, the single photon pulse polarization state dynamics in the fiber should be addressed.

Poon and Law analyzed the polarization and frequency disentanglement of photons via stochastic polarization mode dispersion. They obtained the Schrödingner-like equation and got the master equation by using the Bloch-Redfield-Wangsness approach [9]. Gong et al discussed the dependence of the decoherence of polarization states in phase-damping channels on the frequency spectrum envelope of photons [10]. Zhang and Guo analyzed the decoherence of polarization by quantum master equation and Langevian equation. They obtained that the optical beam is depolarized exponentially [17]. In this letter, we use the birefringence model proposed by Huang and Yevick [11], in which the birefringence vector varies randomly in both magnitude and direction and the amplitude is restricted to a limited range. We focus on the analysis of fidelity of the polarization state and degree of polarization when single photon pulses propagate in a practical optical fiber.

We first give a stochastic polarization mode dispersion model. In this model, optical pulse propagates in a single mode fiber which is a linear and birefringence medium. The propagation direction is along the z-axis, the origin of which is at the input end of fiber. Let fiber birefringence vector be $\vec{b}(\omega,z)$, $\vec{b}(\omega,z) = b_1\vec{e}_1 + b_2\vec{e}_2 + b_3\vec{e}_3$, where $\vec{e}_1, \vec{e}_2, \vec{e}_3$ are unit vectors in the Stoke's space, $\omega$ is frequency, and $z$ is the distance from input end. The birefringence effect on the field can be derived from wave equation, which gives [9][12]

---

[*] Project supported by the National Natural Science Foundation of China (Grant No. 60572147，No.60672119), the 111 Project (B08038) and the State Key Lab. of Integrated Services Networks (ISN 02080002, ISN 090307).

$$\frac{\partial^2}{\partial z^2}\vec{E}(\omega,z)+k^2(\omega,z)\vec{E}(\omega,z)=0 \tag{1}$$

Here, $k^2(\omega,z)=\beta_0^2(\omega,z)\cdot I+\beta_0(\omega,z)\vec{b}(\omega,z)\cdot\vec{\sigma}$, $\beta_0(\omega,z)$ is the propagation constant without birefringence effect, and the vector $\vec{\sigma}=\hat{\sigma}_1\vec{e}_1+\hat{\sigma}_2\vec{e}_2+\hat{\sigma}_3\vec{e}_3$, with $\hat{\sigma}_1=\begin{bmatrix}0 & 1\\ 1 & 0\end{bmatrix}$, $\hat{\sigma}_2=\begin{bmatrix}0 & -i\\ i & 0\end{bmatrix}$, and $\hat{\sigma}_3=\begin{bmatrix}1 & 0\\ 0 & -1\end{bmatrix}$. I is the unit matrix.

Let

$$\vec{E}(\omega,z)=E(\omega)\exp\left[-i\int_0^z\beta_0(\omega,z')dz'\right]\vec{A}(\omega,z) \tag{2}$$

Where, $E(\omega)$ is the amplitude of the field mode, $\vec{A}(\omega,z)$ is the frequency spectrum of the 2-dimensional Jone's vector $\vec{A}(t,z)$.

$$\vec{A}(\omega,z)=\frac{1}{\sqrt{2\pi}}\int\vec{A}(t,z)\exp(i\omega t)dt \tag{3}$$

From Eq. (1), Eq. (2) and Under slowly-varying approximation we obtain

$$\frac{\partial\vec{A}(\omega,z)}{\partial z}=\frac{-i}{2}\left[\vec{b}(\omega,z)\cdot\vec{\sigma}\right]\vec{A}(\omega,z) \tag{4}$$

We assume that the stochastic nature of PMD model is due to the fluctuations of optical axis along the fiber and the frequency dependence of $\vec{b}(\omega,z)$ is determined by fiber material properties only. Based on these assumptions, we have

$$\vec{b}(\omega,z)=f(\omega)\vec{b}(z) \tag{5}$$

Where $f(\omega)=\gamma\omega+\varsigma\omega(\omega-\omega_0)+\ldots$. In non-dispersion medium, only the first term remains, that is $f(\omega)=\gamma\omega$.

Under the condition of linear birefringence, let [13][14]

$$\begin{aligned}\vec{b}(z)&=2b_x(z)\vec{e}_1+2b_y(z)\vec{e}_2+0\cdot\vec{e}_3\\ &=2b_x(z)\vec{e}_1+2b_y(z)\vec{e}_2\end{aligned} \tag{6}$$

We select the model A1 proposed by Huang and Yevick [11], in which $b_x(z)$ and $b_y(z)$ follow the equation below:

$$b_x(z)=\frac{e^{-\lambda_1 z}}{\lambda_2-\lambda_1}\left[\lambda_2 b_x(0)+\int_0^z g_x(z')e^{\lambda_1 z'}dz'\right]+\frac{e^{-\lambda_2 z}}{\lambda_1-\lambda_2}\left[\lambda_1 b_x(0)+\int_0^z g_x(z')e^{\lambda_2 z'}dz'\right] \tag{7}$$

$$b_y(z) = \frac{e^{-\lambda_1 z}}{\lambda_2 - \lambda_1}\left[\lambda_2 b_y(0) + \int_0^z g_y(z')e^{\lambda_1 z'}dz'\right] + \frac{e^{-\lambda_2 z}}{\lambda_1 - \lambda_2}\left[\lambda_1 b_y(0) + \int_0^z g_y(z')e^{\lambda_2 z'}dz'\right] \quad (8)$$

Where, $\lambda_1 = \frac{1-|2\varepsilon-1|}{2L_f(1-\varepsilon)}$, $\lambda_2 = \frac{1+|2\varepsilon-1|}{2L_f(1-\varepsilon)}$, $g_x(z)$ and $g_y(z)$ are White Gaussian Noise with mean zero and variance $\sigma^2$. $\varepsilon$ corresponds to the mean fluctuation magnitude of the stochastic fiber birefringence, $0 < \varepsilon < 1$. $L_f$ denotes the inverse coupling strength of fiber and equals the shortest fiber correlation length when $\varepsilon \to 1$. Here variance $\sigma^2 = \frac{\pi^2}{2\lambda_{rms}L_b^2}$, where

$$\lambda_{rms} = \frac{1}{(\lambda_2 - \lambda_1)^2}\left(\frac{1}{2\lambda_1} + \frac{1}{2\lambda_2} - \frac{2}{\lambda_1 + \lambda_2}\right),$$ $L_b$ is the beat length.

Then the solution to the Eq. (4) is

$$\vec{A}(\omega, z) = \exp\left[-\frac{i}{2}\int_0^z \vec{b}(\omega, z')\cdot\vec{\sigma}dz'\right]\vec{A}(\omega, 0)$$
$$= \exp\left\{-\frac{i\gamma\omega}{2}\int_0^z \begin{bmatrix} 0 & b_x(z') - ib_y(z') \\ b_x(z') + ib_y(z') & 0 \end{bmatrix}dz'\right\}\cdot\vec{A}(\omega, 0) \quad (9)$$

Let $\vec{A}(\omega, z) = \begin{bmatrix} C_0(\omega, z) \\ C_1(\omega, z) \end{bmatrix}$, here $C_0(\omega, z)$ and $C_1(\omega, z)$ are polarization amplitude of electric field. Let $\phi(\omega)$ be the frequency envelope of the input single-photon wave packet, which follows $\int d\omega|\phi(\omega)|^2 = 1$, then the state vector takes the form [9]:

$$|\psi(z)\rangle = \int \phi(\omega)|\omega\rangle \otimes [C_1(\omega, z)|1\rangle + C_0(\omega, z)|0\rangle]d\omega$$
$$= \int \phi(\omega)|\omega\rangle \otimes \vec{A}(\omega, z)d\omega \quad (10)$$

Where $|\omega\rangle$ is the frequency basis vector defined in the birefringence free system ($b(\omega, z) = 0$), and $|1\rangle$ and $|0\rangle$ respectively correspond to vertical and horizontal polarization basis vectors, $|0\rangle = [1\ 0]^T$, $|1\rangle = [0\ 1]^T$. Note that the phase factor $e^{-i\omega t}$ is removed due to the free field evolution.

The density operator $\rho(z)$ can be given as

$$\rho(z) = \overline{|\psi(z)\rangle\langle\psi(z)|} = \iint d\omega d\omega' \phi(\omega)\phi^*(\omega')[|\omega\rangle\langle\omega'|]\otimes\overline{[\vec{A}(\omega, z)\vec{A}^+(\omega', z)]} \quad (11)$$

Here the bar refers to the ensemble average. We compute the partial trace of $\rho(z)$ for the frequency freedom

$$\rho_s(z) = tr_\omega[\rho(z)] = \int d\omega |\phi(\omega)|^2 \overline{[\vec{A}(\omega,z)\vec{A}^+(\omega,z)]} \qquad (12)$$

If the initial polarization is $|A_0\rangle = \alpha|0\rangle + \beta|1\rangle$, in which $|\alpha|^2 + |\beta|^2 = 1$, then input state of single photon pulse in the fiber channel takes the form:

$$|\psi_{in}\rangle = \int \phi(\omega)|\omega\rangle d\omega \otimes |A_0\rangle \qquad (13)$$

So the density operator of input state is

$$\rho_{in} = \int d\omega |\phi(\omega)|^2 |A_0\rangle\langle A_0| \qquad (14)$$

The fidelity between input state and output state is [15]

$$F(z) = tr\sqrt{\rho_{in}^{1/2}\rho_s(z)\rho_{in}^{1/2}} \qquad (15)$$

The degree of polarization (DOP) of the pulse can be given as [17]

$$DOP(z) = \{2tr[\rho_s^2(z)] - 1\}^{1/2} \qquad (16)$$

Let the frequency spectrum envelope of the input pulse to be Gaussian and its frequency envelope $\phi(\omega)$ is

$$\phi(\omega) = \left(\frac{1}{(\Delta\omega)^2 \pi}\right)^{1/4} \exp\left[\frac{-(\omega-\omega_0)^2}{2(\Delta\omega)^2}\right] \qquad (17)$$

Where $\Delta\omega$ denotes the width of the Gaussian envelope and the peak frequency $\omega_0$ is the 1550nm.

We analyze the variation of fidelity with the fiber length under different mean fluctuation magnitudes of the stochastic fiber birefringence $\varepsilon$. Let $L_b = 20m$, $L_f = 12m$, $\alpha = 0.707$, $\Delta\omega = 20GHz$, and let $\varepsilon$ be 0.2, 0.6 and 0.99 respectively. We perform 10000 trials to compute the density operator at each length. Then, fidelity and DOP can be obtained by Eq. (15) and Eq. (16), as shown in Fig. 1 and Fig.2. It is shown that the fidelity of quantum state decreases quickly and tends to a stable value along optical fiber, and the fidelity increases for larger $\varepsilon$. As shown in Fig.2, the DOP is nearly constant when $\varepsilon = 0.2$, and decreases for larger $\varepsilon$.

Besides Gaussian frequency spectrum envelope we also analyze the fidelity and DOP for Lorentzian and rectangular spectrum. Lorentzian frequency spectrum is given by [10]

$$\phi(\omega) = \left(\frac{\Delta\omega}{\pi}\right)^{1/2} \frac{1}{\sqrt{(\Delta\omega)^2 + (\omega-\omega_0)^2}} \qquad (18)$$

The rectangular spectrum is given by [10]

$$\phi(\omega) = \begin{cases} \frac{1}{\sqrt{2\Delta\omega}} & |\omega-\omega_0| \leq \Delta\omega \\ 0 & |\omega-\omega_0| > \Delta\omega \end{cases} \qquad (19)$$

The fidelities of quantum states and DOP of the pulse under these three kinds of frequency spectrum envelopes are shown in Fig.3 and Fig. 4. Let $L_b = 20m$, $L_f = 12m$, $\alpha = 0.707$, $\Delta\omega = 20GHz$,

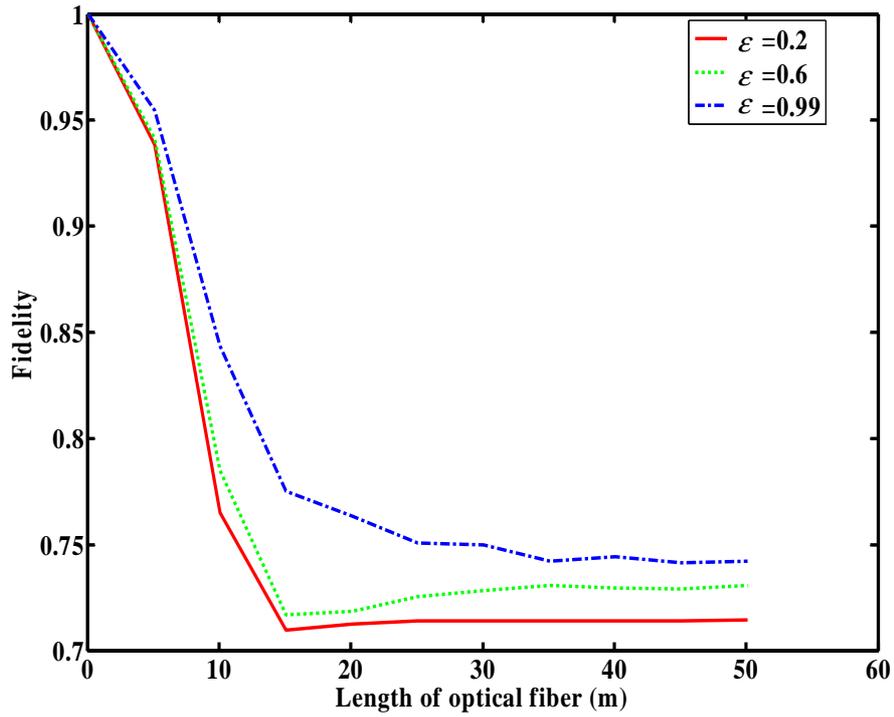

Fig. 1 Fidelities with different mean fluctuation magnitudes
of the stochastic fiber birefringence

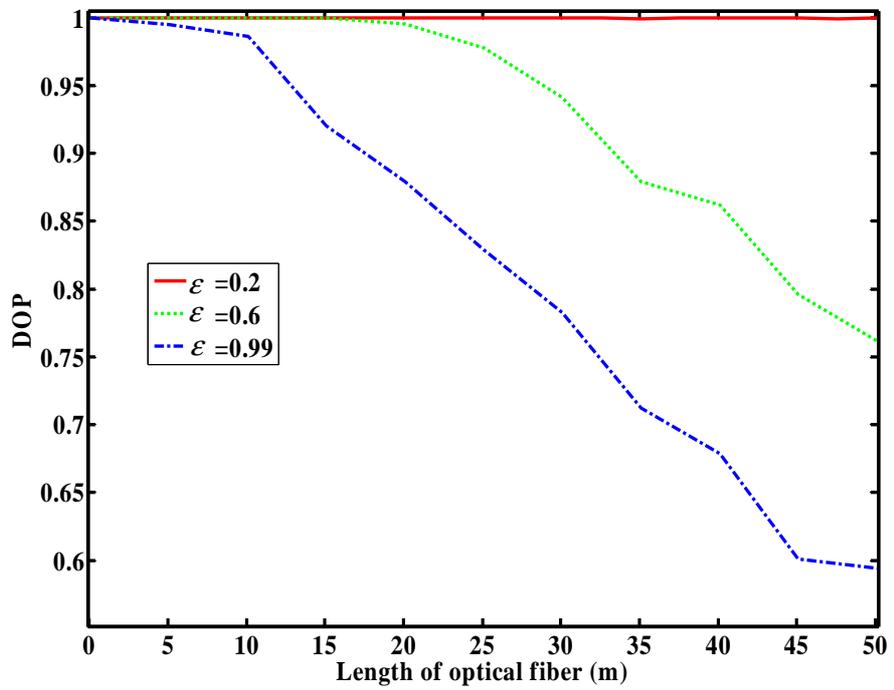

Fig. 2 Degree of polarization with different mean fluctuation magnitudes
of the stochastic fiber birefringence

$\varepsilon = 0.6$. The fidelities vary in the same way for Gaussian and rectangular frequency spectrum envelope, while the value of Lorentzian spectrum is smaller. The DOP vary nearly in the same way for Gaussian and rectangular frequency spectrum envelope, while the value of Lorentzian spectrum is smaller.

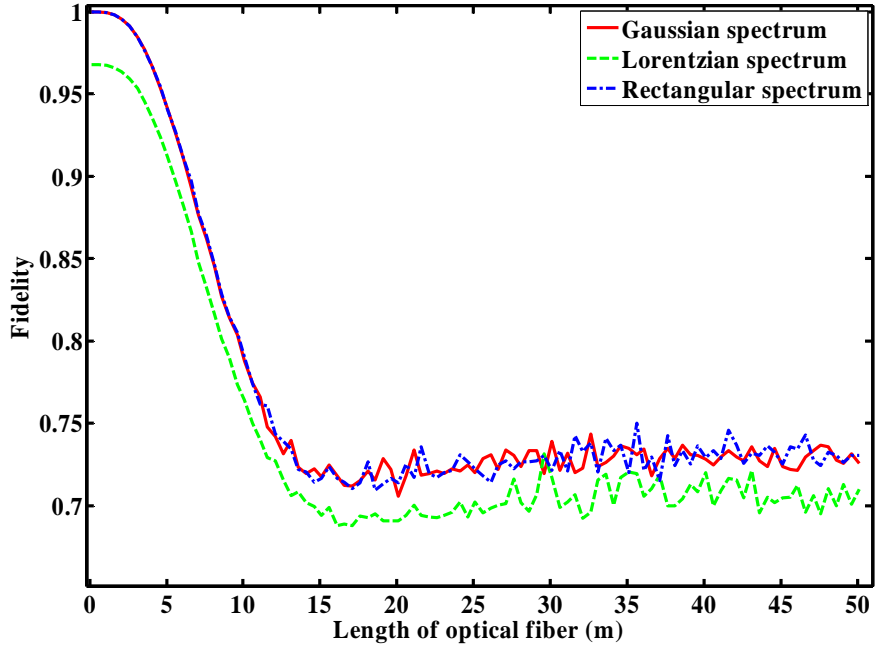

Fig.3 The fidelities of quantum states under Gaussian,
Lorentzian and rectangular frequency envelope

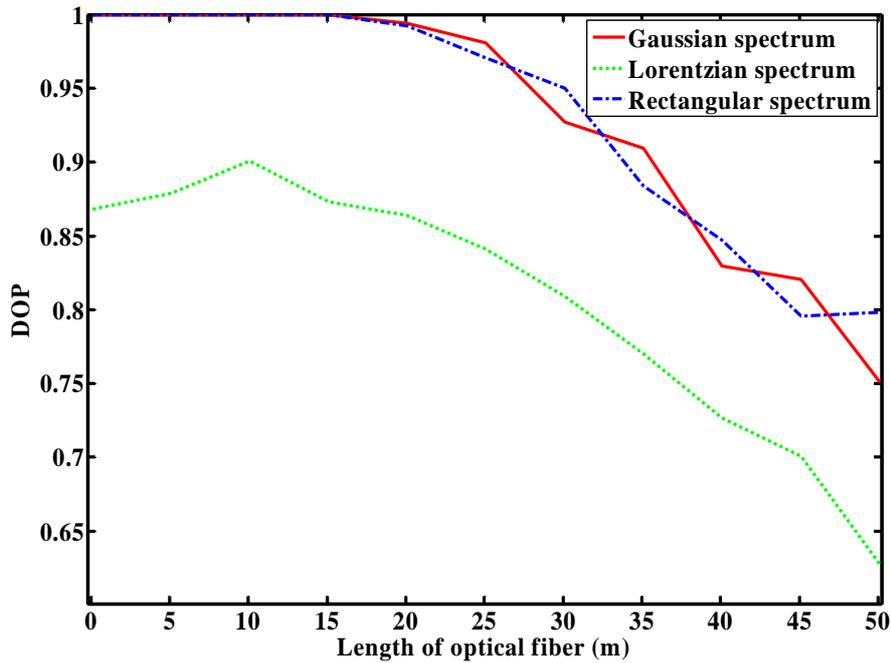

Fig.4 The variation of degree of polarization with the fiber length under
Gaussian, Lorentzian and rectangular frequency envelope

In conclusion, we have discussed the polarization state dynamics of single photon pulse propagating in an optical fiber with stochastic polarization mode dispersion based on a general birefringence vector model. We can predict the possible variation of polarization state while a pulse propagates in a fiber if we can estimate the channel parameters previously. Furthermore, we can implement effective dynamic polarization control. In this letter, insert loss of optical fiber, frequency dispersion, and non-linear effect [16], etc. are not taken into account. The application of the results in practical quantum key distribution system need further discuss and practice.